\documentclass[12pt,preprint]{aastex}

\begin{document}

\title{ON THE ORIGIN OF LYMAN-$\alpha$ BLOBS AT HIGH REDSHIFT:
       KINEMATIC EVIDENCE FOR A HYPERWIND GALAXY AT $z$ = 3.1
\altaffilmark{1}}

\altaffiltext{1}{Based on data collected at the Subaru Telescope, which is
       operated by the National Astronomical Observatory of Japan.}

\author{Youichi Ohyama\altaffilmark{2}, Yoshiaki Taniguchi\altaffilmark{3}, Koji S. Kawabata\altaffilmark{4},
Yasuhiro Shioya\altaffilmark{3}, Takashi Murayama\altaffilmark{3}, Tohru Nagao\altaffilmark{3},
Tadafumi Takata\altaffilmark{2}, Masanori Iye\altaffilmark{4}\altaffilmark{5},
and Michitoshi Yoshida\altaffilmark{6}}

\altaffiltext{2}{Subaru Telescope, National Astronomical Observatory
            of Japan, 650 N. A`ohoku Place, Hilo, HI 96720}

\altaffiltext{3}{Astronomical Institute, Graduate School of Science,
           Tohoku University, Aramaki, Aoba, Sendai 980-8578, Japan}

\altaffiltext{4}{National Astronomical Observatory of Japan, 2-21-1 Osawa,
           Mitaka, Tokyo 181-8588, Japan}

\altaffiltext{5}{Department of Astronomy, Graduate University for Advanced
                Studies, 2-21-1, Osawa, Mitaka, Tokyo 181-8588, Japan}

\altaffiltext{6}{Okayama Astrophysical Observatory, National Astronomical
               Observatory of Japan, Kamogata, Asakuchi, Okayama
               719-0232, Japan}

\begin{abstract}
We present deep optical spectroscopy of an extended Ly$\alpha$
emission-line blob located in an over-dense region at redshift $z
\approx 3.1$; `blob 1' of Steidel et al. (2000). The origin of
such Ly$\alpha$ blobs has been debated for some time; two of the
most plausible models are (1) that it comes from a
dust-enshrouded, extreme starburst galaxy with a large-scale
galactic outflow (superwind/hyperwind) or (2) that it is the
cooling radiation of proto-galaxies in dark matter halos.
Examination of the kinematic properties of the Ly$\alpha$
emission-line gas should allow us to determine its nature. With
this motivation, we performed optical spectroscopy of `blob
1' using the Subaru Telescope, and found that its kinematic
properties can be well explained in terms of superwind activity.
\end{abstract}


\section{INTRODUCTION}

Steidel et al. (2000, hereafter S00) found two unusual Ly$\alpha$
emission-line objects in an over-dense region at redshift $z
\approx 3.09$. These two objects are known as Ly$\alpha$
emission-line blobs (hereafter LABs).
These two LABs comprise the most remarkable class of
high-redshift objects observed to date because they are strikingly
larger than typical high-redshift Ly$\alpha$ emitters. Their
observational properties are summarized below \footnote{ We adopt
a flat-universe cosmology, with $\Omega_{\rm matter} = 0.3$,
$\Omega_{\Lambda} = 0.7$, and $h=0.7$ where $h = H_0/($100 km
s$^{-1}$ Mpc$^{-1}$), throughout this {\it Letter}. In this
cosmology, the angular size distance to blob 1 at $z=3.09$ is
$D_{\rm A} = 1.58$ Gpc, so that 1\arcsec\ corresponds to 7.63 kpc.
Note also that its luminosity distance is $D_{\rm L} = 26.3$ Gpc.
}:
(1)the observed Ly$\alpha$ luminosities are $\sim 10^{43}$
ergs s$^{-1}$;
(2) they are resolved and elongated;
(3) their sizes are $\sim$ 100 kpc,
(4) the observed line widths (full width at zero intensity;
see Fig. 8 of S00) are $\sim 1000$
km s$^{-1}$, and
(5) they are not associated with a strong radio continuum source
such as a powerful radio galaxy or a quasar; i.e., there is no evidence
for an active galactic nucleus (AGN).

Two models have been suggested for the emissions from these LABs.
The first is that they are superwinds/hyperwinds driven by the
initial starbursts in forming galaxies; all of the above
observational properties, as well as the observed frequency of
LABs, can be explained in terms of this superwind model (Taniguchi
\& Shioya 2000; hereafter TS00). TS00 also discuss the
evolutionary link between dust-enshrouded (or dusty) submillimeter
sources and LABs; they argue that the central
starburst region in a forming elliptical galaxy would be
enshrouded by large amounts of gas and dust.
This superwind model predicts that the LABs should be bright in
the rest-frame far-infrared, and thus at observed submillimeter
wavelengths. Indeed, strong submillimeter continuum emission has
been detected in the blob 1 (hereafter LAB1; Chapman et al. 2001).
Taniguchi, Shioya, \& Kakazu (2001; hereafter TSK01) 
investigated the spectral energy distribution (SED) of LAB1,
and found it quite similar to that of Arp 220.
They also found that its infrared luminosity exceeds $10^{13}
L_\odot$, and therefore proposed that LAB1 is a hyperluminous
infrared galaxy or a hyperwind galaxy at $z = 3.1$.

The second model suggests that LABs are cooling radiation from
proto-galaxies within dark matter halos (Haiman, Spaans, \& Quataert
2000; Fardal et al. 2001; Fabian et al. 1986; Hu 1992). Standard cold
dark matter models predict that a large number of dark matter halos
collapse at high redshift; proto-galaxies in the halos can emit
significant Ly$\alpha$ fluxes through collisional excitation of
hydrogen. The expected properties of these Ly$\alpha$-emitting halos
are also consistent with the observed linear sizes, velocity widths,
and Ly$\alpha$ fluxes of the LABs. However, it is uncertain how much
far-infrared and submillimeter continuum emission should be emitted in
this model, because little is known about the dust content and its
spatial distribution in such dark matter halos.

One of the key ways to discriminate between these two models is to
investigate the kinematical properties of the Ly$\alpha$-emitting
nebula in detail; the mechanism of the velocity broadening of the
emission lines is different in the two models. S00 presented
optical spectroscopy of the two LABs and showed that the
Ly$\alpha$ nebulae have peculiar velocity fields with FWHM (full
width at half maximum) up to $\simeq$ 2000 km s$^{-1}$. However,
since their slits did not cover the central regions of the LABs,
their results could not be used to distinguish between the two
models. In this Letter, we present new, deep optical spectroscopy
for LAB1 using the Subaru Telescope.

\section{OBSERVATIONS}

Spectroscopic observations were made with the FOCAS
(Kashikawa et al. 2002) on the
Subaru Telescope (Kaifu 1998) on 17 June 2002 (UT). The seeing was
0\farcs5 during the observations. To observe both the line and
continuum emission properties of LAB1, we aligned the
1\farcs0-wide slit so as to observe both the Ly$\alpha$ peak of
LAB1 and the $K$ source found by S00 at a position angle of
134$^\circ$. The 300B grism together with a Y47 filter allowed us
to obtain an optical spectrum between 4700 \AA ~ and 9400 \AA ~
with a spectroscopic resolution of 18 \AA ~ (the instrumental
FWHM) at $\lambda$5577 \AA, or 970 km s$^{-1}$. The pixel sampling
was 0\farcs4 in angle (4-pixel binning) and 5.65 \AA ~ in
wavelength (4-pixel binning). We obtained seven spectra, each one
based on an exposure of 1800 s. An offset of 2\farcs0 along the
slit was applied between adjacent exposures; i.e., we operated in
a nodding mode with $\pm$ 2\arcsec\ offsets. The total exposure
time was 3.5 h.
Data reduction was made in a standard manner (bias subtraction,
flat fielding, optical distortion corrections, wavelength calibration,
and flux caribration) using specialized software developed by the FOCAS
team (Yoshida et al. 2000) as well as IRAF.

We imaged the source on the same night in $R$ band using the direct
imaging mode of FOCAS. We obtained two $R$-band images with a CCD
binning of $2\times 2$ pixels ($0.2\times 0.2$ arcsec$^2$ per binned
pixel). Each exposure time was 300 s, so that the total
integration was 600 s. The sky conditions were photometric.
Bias subtraction, flat-fielding, and optical distortion corrections were
applied using the FOCAS software and IRAF.
The flux calibration was made using flux conversion data obtained on
another photometric night with the same observing settings.

\section{RESULTS}

\subsection{Kinematical Properties}

In Fig. 1, we show the slit position for our optical spectroscopy
overlaid on
(1) Ly$\alpha$, (2) $K$-band, (3) $R$-band images, and (4) a zoomed $K$-band
image overlaid on the $R$-band image.
Ly$\alpha$ and $K$-band images are taken from S00.
The $R$- and
$K$-band images were registered by utilizing the positions of four
brighter objects present in both images; the positional error of
registration is estimated to be $\simeq 0.5$''. We also show our
Ly$\alpha$ spectrogram as a greyscale in the first panel. Our
slit covers the peak of Ly$\alpha$ emission and the $K$ source.

In Fig. 2, we show a one-dimensional spectrum over a wider wavelength range
(4700\AA -- 8200\AA) extracted from the central 1\farcs2 of the source.
To show the detailed kinematical properties of the nebula, we also show
our Ly$\alpha$ spectrogram together with the
peak-normalized spectrogram.
In the latter panel, the intensity peak of the profiles is normalized to be
unity along the spatial direction, so that it is possible to see interesting
velocity structures at fainter intensity levels.
The peak-normalized sky emission line at 5577 \AA ~ is
also shown at the same intensity scale as that of the peak-normalized
Ly$\alpha$ spectrogram to illustrate the instrumental velocity resolution.
It can be seen from this Figure that the inner part of the nebula
($<3$\arcsec\ from the center) is clearly resolved in velocity.
We have performed single and/or multiple Gaussian component fittings
to measure the kinematical properties of the Ly$\alpha$ nebula.
In the central 2\arcsec\ region, we find the following
three emission-line components (see also Fig. 3):
(1) The main component is centered
at $\lambda \approx 4981$ \AA, giving a redshift of $z = 3.097 \pm
0.002$. Its FWHM\footnote{All the line widths in this Letter are corrected for
the instrumental resolution of 970 km s$^{-1}$ FWHM.}
is $\simeq 1500$ km s$^{-1}$.
(2) An additional redshifted component is seen; its peak velocity is
shifted by $\simeq +$2500 km s$^{-1}$ with respect to that of the main
component. Its
FWHM is measured to be $\simeq 1000$ km s$^{-1}$.
(3) A blueshifted
component is also seen; its peak velocity is shifted by
$\simeq -$3000 km s$^{-1}$ with respect to that of the main component. Its
FWHM ranges from 2500 km s$^{-1}$ to 3500 km s$^{-1}$.
The presence of this blueshifted component makes the global Ly$\alpha$
emission-line profile blueward asymmetric (see middle panel of Fig. 3).
Double-peaked profiles are found both at 1-2\arcsec\ NW of the center
and at 1\farcs5 - 2\farcs5 SE of the center. Although the double peaks
are not clearly separated in the spectrogram, the profiles seem to be
well deconvolved by the double Gaussian components (Fig. 3) since the
total line width is wider than those of the outer regions (Fig. 2) and
the profiles show a flat-topped shape (Fig. 3). The velocity
difference between the two peaks is $\Delta v = v_{\rm red} - v_{\rm
blue} \simeq$ 1200 km s$^{-1}$. Outside these double-peaked profile
regions (at radial distances from the center $r \simeq 3$ - 5''),
we see that the emission line profile is singly peaked on
either side. The FWHM in these regions ranges from 1000 km s$^{-1}$ to
1500 km s$^{-1}$. We find no evidence for any global rotation within
$r < 5$''.
All these observed kinematical
properties are summarized in Fig. 4.

\subsection{The $R$-band Counterpart of the $K$ Source}

Our observations detect a probable $R$-band counterpart of the
$K$-band source. S00 found the $K$ source ($K_{\rm s} \simeq$
21.2) close to the peak of both the Ly$\alpha$ emission line and
the submillimeter peak, and thus suggested that it was probably
the host galaxy of LAB1. Although they could not detect its $R$-band
counterpart ($R > 25.7$),
we have detected a probable $R$-band counterpart of LAB1 (hereafter the $R$
source) at $R=26.1\pm 0.3$ for the first time.
The $R$ source is found at $\approx$
1\arcsec\ south west of the $K$ source, not at the same position.
It has an amorphous shape without any obvious light concentration,
and appears to surround the $K$ source along its south-western
side. Although such a distinctive shape suggests a physical
association between the $R$ and $K$ sources, its origin is
unknown, since our slit was not placed on the $R$ source due to
its slight displacement (Fig. 1).

\section{DISCUSSION}

One of possible origin of LAB1 is scattering of the nuclear
Ly$\alpha$ photons within the halo. However, since the Ly$\alpha$
line is a resonance line and a multiple scattering process is
required to reproduce such a huge nebula ($\simeq 100$ kpc) in the
Case B condition (Osterbrock 1989), the Ly$\alpha$ flux would be
expected to be reduced significantly by dust grains within the
halo (e.g., Ferland \& Netzer 1979). In the case of LAB1, there
are some lines of evidence for the presence of a significant
amount of dust, such as the extremely high FIR luminosity
($>10^{13}$ $L_{\rm \odot}$), most probably originating from warm
dust (TSK01), and the red color of the $K$ source ($R-K>4.5$)
which is $\simeq 2$ magnitude redder than that of a nearby Lyman
break galaxy, C11, at the same redshift (see Fig. 7 in S00).
Therefore, scattering of a nuclear source does not seem a likely
explanation for the observations of LAB1.

Another possible explanation of the Ly$\alpha$ emission from the LABs
is the cooling radiation from proto-galaxies within dark matter halos
(Haiman et al. 2000; Fardal et al. 2001; Fabian et al. 1986; Hu 1992).
This model may also explain the observed velocity widths as well as
the linear sizes and the Ly$\alpha$ fluxes of LABs (Haiman et al.
2000). The wider velocity widths of the Ly$\alpha$ lines from proto-galaxies
are attributed to the frequency shift of the Ly$\alpha$ emission
required for the photons to escape from the optically thick core of a
Voigt profile (i.e., diffusion). Therefore, this model predicts that
only simple line profiles should be observed over the halo. However,
our optical spectroscopy reveals
a complex, structured velocity field.
These observed properties do not have a straightforward explanation in
the cooling-radiation model. Since the superwind model
does not have difficulty in explaining these properties (see
below), our observations favor the superwind model over the cooling
radiation model.

Possible alternative ideas for explaining LAB1 are an inflow of halo
material around the young, forming galaxy or an outflow of the
superwind associated with star-formation activity of the host galaxy.
Detection of double-peaked profiles on large scales around the central
source ($r \simeq 1$ - 2'', or $\simeq$ 7.6 - 15 kpc) provides
good constraints on possible models, since this type of profile
requires an inflowing or outflowing shell (or cone) of material around
the galaxy. The presence of such an expanding shell or cone is a
natural consequence of superwind activity (e.g., Heckman et al. 1990),
whereas the inflow of halo matter would not necessarily take the form
of a large-scale shell. Other pieces of kinematical evidence also
favor the model in which the outflow is associated with a superwind.
The overall blueward asymmetric profile in the central region is often
taken to be a signature of superwind galaxies (e.g., Taniguchi et al.
1988; Heckman, Armus, \& Miley 1990). The presence of both the
blueshifted ($-$3000 km s$^{-1}$) and the redshifted ($+$2000 km
s$^{-1}$) components in the central region can be attributed to the
expanding motion of the shocked shell (see TS00). The main component
is considered to be the central starburst region, which has the
systemic velocity of the host galaxy. Although the double-peaked
profiles disappear further out ($r = 3 - 5$'', or $23 -
38$ kpc), this can be understood if the peaks' velocity separation is
narrower than our velocity resolution (i.e., 970 km s$^{-1}$). This
would come about if the wind had a slower speed in the outer regions
due to momentum loss as it interacted with the ambient halo gas.
Therefore the superwind/starburst scenario for the LAB1 looks more
plausible than other models, from a kinematical point of view. In this
model, the velocity gradient within the nebula (the velocity
splitting) would help the Ly$\alpha$ photons generated by the local
shock within the nebula to escape, although a fraction of the photons
would still be absorbed by dust during their scattering within the nebula.
Although the superwind model suggests that emission-line
morphology should appear either as a bi-conical nebula or as a pair of
superbubbles (e.g., Heckman et al. 1990), the LAB1 appears chaotic
probably due to a strong interaction between
the superwind and ambient gaseous matter around the host galaxy.

In conclusion, all the kinematical properties of LAB1 can be
explained by the superwind model proposed by TS00. Together with
the submillimetric evidence presented by TSK01, we conclude that
LAB1 is one of the largest superwind (or hyperwind; TSK01)
galaxies in the early universe observed to date. Such superwind
phenomena are a natural consequence of the intense starbursts that
are expected to occur in forming galaxies (Larson 1974; Arimoto \&
Yoshii 1987).
Indeed, recent optical spectroscopy has found
possible superwind galaxies beyond $z=5$ (Dawson et al. 2002;
Ajiki et al. 2002).
Since the surface brightness of LABs is very
low, future deep imaging surveys with narrow-band filters will be
necessary to detect larger numbers of LABs in the early universe.

\acknowledgments
YS was a JSPS fellow. This work was financially supported in part
by the Ministry of Education, Science, and Culture, Japan (Nos.
10044052, and 10304013).

%\newpage
%----------------------------------------------------------------------------
%      References
%----------------------------------------------------------------------------

\newpage

\begin{figure}
\epsscale{0.8}
\plotone{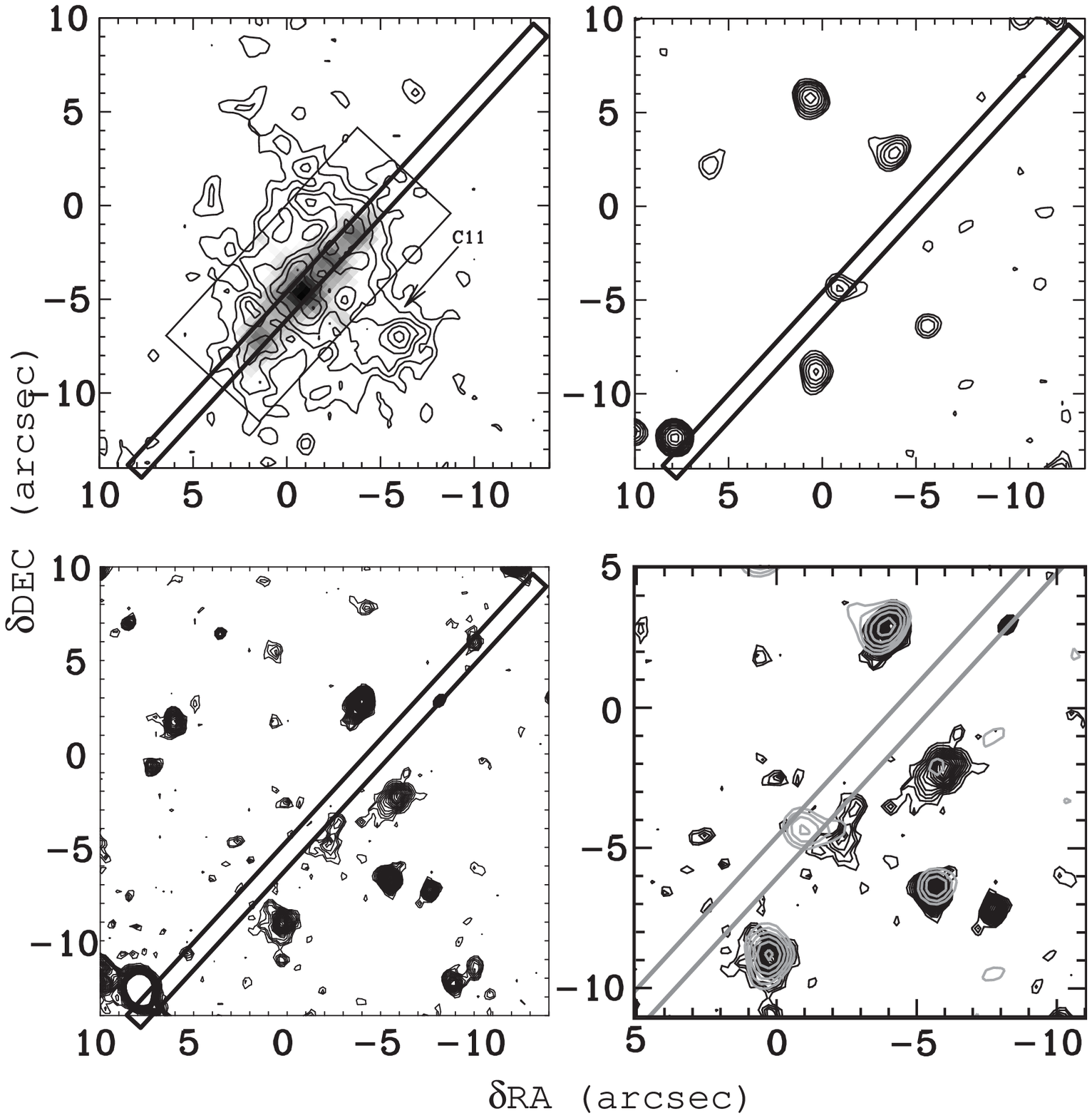}
\figcaption{ The slit position used in our optical spectroscopy is
overlaid on (1) the Ly$\alpha$ image (contours) taken from S00
(upper left panel), (2) the $K$-band image taken from S00
(upper right panel), (3) the $R$-band image taken in our
observations (lower left panel), and (4) an expanded $K$-band
image overlaid on the $R$-band image (lower right panel). The $K$-
and $R$-band contours are shown in light gray and black,
respectively, in the lower right panel. North is up and east is to
the left in all panels. We also show our Ly$\alpha$ spectrogram in
the upper left panel as a greyscale.}
\end{figure}

\begin{figure}
\plotone{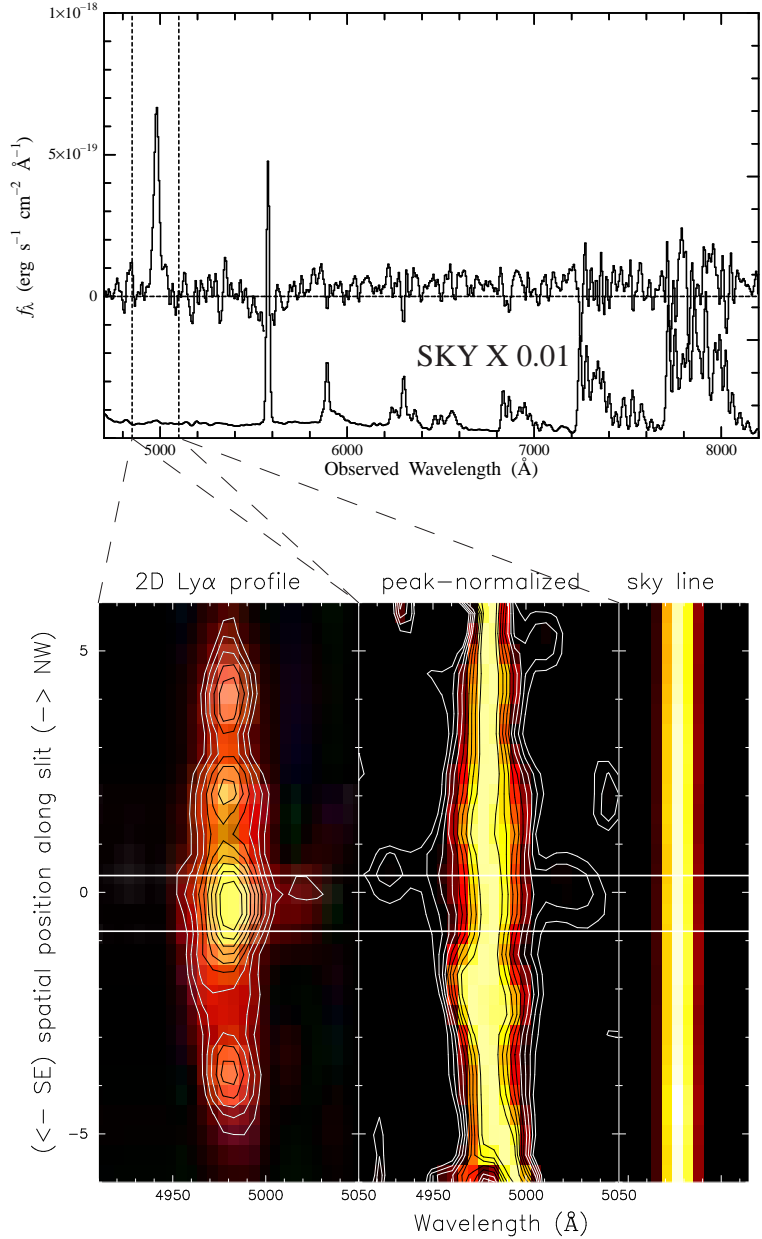}
\figcaption{ ($top$) A one-dimensional spectrum over a wider wavelength
range (4700\AA -- 8200\AA) extracted from a central 1.2''-wide region.
A one-dimensional sky spectrum extracted in the same way is also shown
on a reduced scale ($\times 0.01$). ($bottom$) The observed Ly$\alpha$
spectrogram (left panel) is shown together with the peak-normalized
spectrogram (middle panel) and the peak-normalized sky emission line
at 5577 \AA ~ (right panel). The middle and right panels can be
compared directly to show that the Ly$\alpha$ emission is resolved in
width. Contours are drawn at 3$\sigma$, 4$\sigma$, ..., 12$\sigma$ of
the background noise level in the left panel, and are 10\%, 20\%, ...,
90\% of the peak flux in the middle panel. Two horizontal lines show
the region of extraction of the one-dimensional spectrum shown above.}
\end{figure}

\begin{figure}
\plotone{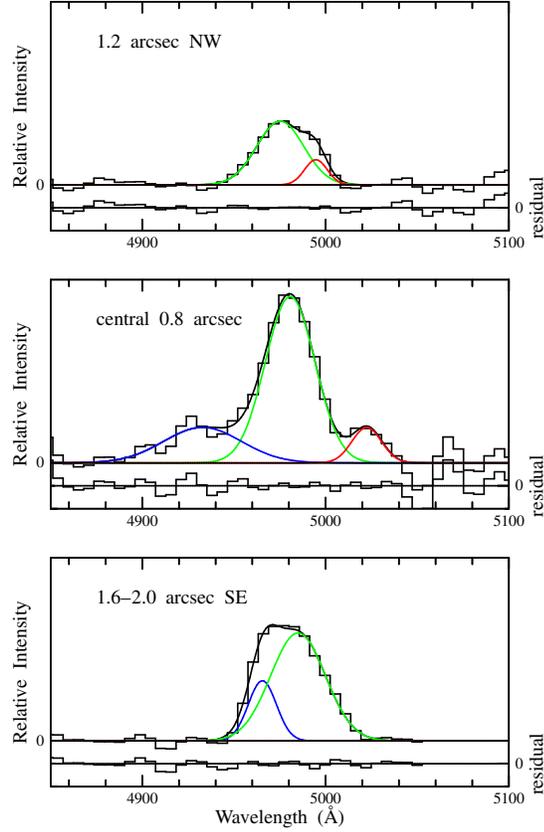}
\figcaption{ Emission-line profiles at various positions; (a)
$+1\farcs2$ NW of the Ly$\alpha$ peak, (b) the central 0\farcs8 region,
and (c) 1\farcs6 - 2\farcs0 SE of the Ly$\alpha$ peak.
Two-component Gaussian fitting has been applied to cases (a) and
(c) while a three-component fit was carried out for case (b). The
fitting results for the main component are shown by the green
curves. Those for the blueshifted and the redshifted components
are shown by the blue and red curves, respectively. The fitting
residuals are shown at the bottom of each panel.}
\end{figure}

\begin{figure}
\plotone{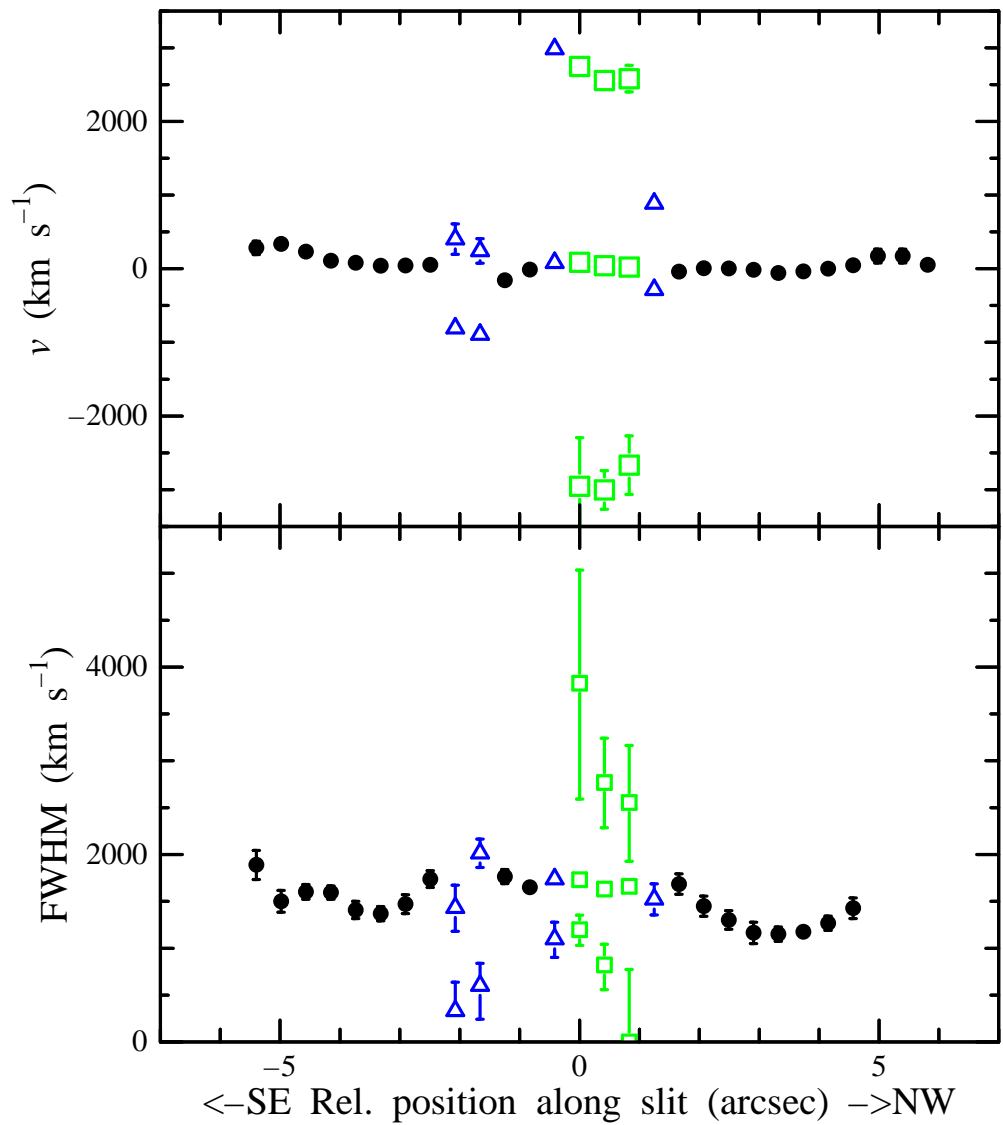}
\figcaption{
The spatial variations of the peak velocities of the Ly$\alpha$ nebula
(upper panel) and the FWHMs corrected for the instrumental resolution
(lower panel). Data for the triple-peaked
regions and the double-peaked regions are shown by open squares
and open triangles, respectively. Those for the single-peaked regions
are shown by filled circles. Measurement errors are shown by vertical
bars.}
\end{figure}
\end{document}